\documentclass[floatfix,reprint,aps,prb,twocolumn,amssymb,superscriptaddress,showkeys]{revtex4-2}

\usepackage{mathtools} 
\usepackage{graphicx}
\usepackage{epstopdf}
\usepackage{amsmath}
\usepackage{dcolumn}
\usepackage{bm}
\usepackage{bbold}
\usepackage{amssymb}
\usepackage{mathrsfs}
\usepackage{amsfonts}
\usepackage{braket}
\usepackage{tabularx}
\usepackage{cleveref}
\usepackage{setspace}
\usepackage[caption=false]{subfig}

\usepackage{float}
\usepackage[toc,page]{appendix}
\usepackage{array}
\usepackage[dvipsnames]{xcolor}
\newcolumntype{L}[1]{>{\raggedright\let\newline\\\arraybackslash\hspace{0pt}}m{#1}}
\newcolumntype{C}[1]{>{\centering\let\newline\\\arraybackslash\hspace{0pt}}m{#1}}
\newcolumntype{R}[1]{>{\raggedleft\let\newline\\\arraybackslash\hspace{0pt}}m{#1}}

\newcommand{\figref}[1]{Fig.~\ref{#1}}

\begin{document}


\title{Strong to ultra-strong coherent coupling measurements in a YIG/cavity system at room temperature}

\author{Guillaume Bourcin}
\email{guillaume.bourcin@imt-atlantique.fr}
\affiliation{IMT Atlantique, Technopole Brest-Iroise, CS 83818, 29238 Brest Cedex 3, France}
\affiliation{Lab-STICC (UMR 6285), CNRS, Technopole Brest-Iroise, CS 83818, 29238 Brest Cedex 3, France}
\author{Jeremy Bourhill}
\affiliation{Quantum Technologies and Dark Matter Labs, Department of Physics,  University of Western Australia, 35 Stirling Hwy, 6009 Crawley, Western Australia.\\
}%
\author{Vincent Vlaminck}
\affiliation{IMT Atlantique, Technopole Brest-Iroise, CS 83818, 29238 Brest Cedex 3, France}
\affiliation{Lab-STICC (UMR 6285), CNRS, Technopole Brest-Iroise, CS 83818, 29238 Brest Cedex 3, France}
\author{Vincent Castel}
\email{vincent.castel@imt-atlantique.fr}
\affiliation{IMT Atlantique, Technopole Brest-Iroise, CS 83818, 29238 Brest Cedex 3, France}
\affiliation{Lab-STICC (UMR 6285), CNRS, Technopole Brest-Iroise, CS 83818, 29238 Brest Cedex 3, France}

\date{\today}

\begin{abstract}
    We present an experimental study of the strong to ultra-strong coupling regimes at room temperature in frequency-reconfigurable 3D re-entrant cavities coupled with a YIG slab. The observed coupling rate, defined as the ratio of the coupling strength to the cavity frequency of interest, ranges from 12\% to 59\%. We show that certain considerations must be taken into account when analyzing the polaritonic branches of a cavity spintronic device where the RF field is highly focused in the magnetic material. Our observations are in excellent agreement with electromagnetic finite element simulations in the frequency domain.

    %
\end{abstract}


\maketitle

\section{Introduction}
Cavity spintronics is an emerging research field that investigates light-matter interactions within magnetism, specifically the interactions between cavity photons and the quanta of spin waves based on the magnetic dipole interaction {--} magnons. At the core of cavity spintronics are cavity-magnon polaritons (CMPs) which are the associated bosonic quasiparticles, i.e., hybridized cavity-magnon-photon states in the strong coupling regime. cavity spintronics has drawn a growing interest since the first theoretical prediction in 2010 \citep{PhysRevLett.104.077202}, and then  shortly after experimental demonstration of CMPs at both millikelvin (mK) temperatures \citep{Huebl2013, Tabuchi2014} and room temperature (RT) \citep{Bai2015}. Cavity spintronics display a broad range of applicability for quantum information systems and RF devices such as adjustable sensitive filter \citep{HarderDissipative, Lachance, ReviewCavMag}, isolators or circulators \citep{Fuller1987}, gradient memories \citep{Zhang2015} and for engineering chiral states of electromagnetic radiation \citep{Bourhill2023, Yu2020}.

In a cavity--magnon system, when the magnon frequency is tuned by an externally applied static magnetic field towards the cavity resonance frequency, the system undergoes hybridization (e.g. forms a CMP) with a characteristic anti-crossing signature in the dispersion spectrum. The interaction is quantified by the coupling strength $g/2\pi$ and by its ratio $g/\omega$ with the cavity frequency $\omega/2\pi$. When the coupling $g/2\pi$ is larger than the systems losses, there exist three different coupling regimes. These have commonly been referred to as: (i) Strong coupling when $g/\omega<0.1$, (ii) Ultra-Strong Coupling (USC) for $0.1<g/\omega<1$, and (iii) Deep-Strong Coupling (DSC) for $g/\omega>1$, a regime that still remains largely unexplored. The value of $g/\omega=0.1$ is considered as a threshold between the SC and USC regimes, but this is only a historical convention, supposedly indicating the cutoff beyond which the coupling rate $g$ represents a {``}sizeable fraction{''} of the system energy and therefore cannot be deemed to be a slowly rotating term in the rotating wave approximation.

The USC regime was predicted theoretically in intersubband cavity polaritons in 2005 \citep{USCprediction} and first observed in 2009 \citep{USCfirstEXP} in n-doped GaAs quantum wells embedded in a microcavity, with $g/\omega=0.11$. Since this experimental observation, several research groups have experimentally achieved the USC regime \citep{ReviewUSC,ReviewUSC2} in different systems such as superconducting circuits \citep{superconductingUSC}, polaritons \citep{landauUSC}, and optomechanics \citep{optoUSC}. So far, the USC regime in cavity spintronics has been experimentally achieved at low temperature \citep{USC-magnonphoton-10, USC-magnonphoton-7, USC-magnonphoton-22, USC-magnonphoton-46, USC-magnonphoton-24, ApproachDSC, ApproachDSCbis} and investigated theoretically \citep{USCtheoryGerrit01, USCtheoryGerrit02}.

Very recently, Golovchanskiy et al. \citep{ApproachDSCbis} proposed an approach to achieve on-chip USC hybrid magnonic systems  reaching $g/\omega=0.6$ and based on superconducting/insulating/ferromagnetic multilayered microstructures operating below 10 K. They highlighted in particular the drastic failure of currently adopted models in the USC regime.

Here, we present measurements and simulations of a reconfigurable hybrid system that allows the study of the transition from the SC to USC regimes at room temperature in the $0.1-15$ GHz frequency range. We utilize a magnetic field-focusing double-post re-entrant cavity first described by Goryachev et al. \citep{USC-magnonphoton-10}.
A set of three different resonators (by their dimensions and posts shape) allow us to follow the evolution of the coupling strength through USC regime (starting from the SC/USC limit). With these results, we confirm that it is necessary to add an extra term in the expression of the Ferromagnetic Resonance (FMR) frequency equation to accurately describe the observed hybridization (measurements and simulations) with the commonly used Dicke model \citep{Dicke1954}. We show that this additional term does not depend on the coupling rate but on the level of confinement of the RF magnetic field in the magnetic material. Moreover, this added term can be negligeable in the SC regime, while it is essential in the USC regime.

\section{Hybrid system description}

The hybrid system presented here is made of a commercial single crystal of YIG (Yttrium Iron Garnet, Y$_3$Fe$_5$O$_{12}$) and a modified re-entrant cavity. The YIG is a slab of 3.82$\times$6.09$\times$0.61 mm$^3$.

\begin{figure}[ht!]
    \centering
    \includegraphics[width=\linewidth]{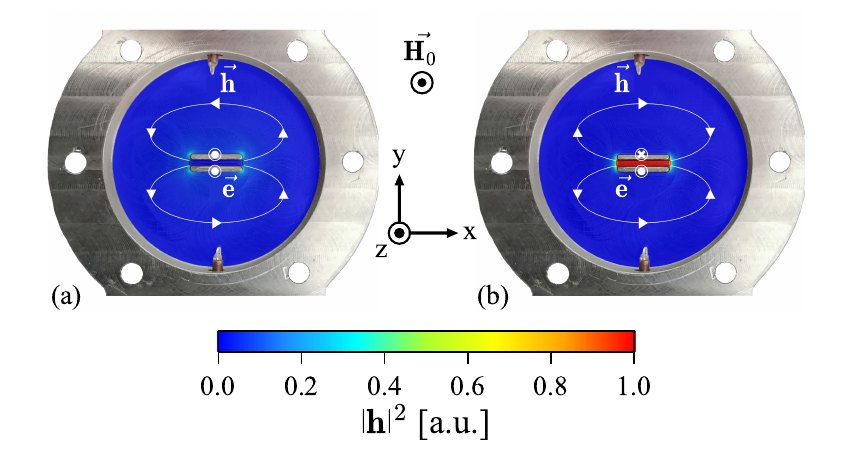}
    \caption{Re-entrant cavity with electromagnetic simulation overlay where $|\mathbf{h}|^2$ is displayed for the first two photonic modes: (a) the DM; and (b) the BM.}
    \label{fig:hybrid}
\end{figure}

The multiple post re-entrant cavity \citep{USC-magnonphoton-10} is a unique type of microwave cavity. There are two first-order resonant modes, termed the Dark Mode (DM) and the Bright Mode (BM). Both contain the electric field of the mode between the top of the post and the lid of the cavity. For the DM (as shown in \figref{fig:hybrid} (a)), the RF electric fields ($\mathbf{e}$-fields) focused above the two posts are in-phase, resulting in the circulating RF magnetic fields ($\mathbf{h}$-fields) destructively interfering in the region between the posts (hence {``}dark{''}), whilst the opposite is true for the BM (as shown in \figref{fig:hybrid} (b)). The advantages of such a cavity are three-fold: first, the highly localized electric field results in extremely large frequency sensitivity to any perturbations inside this region (displacement of the containment area or modification of the dielectric material). Secondly, the physical separation of the electric and magnetic fields permits separate interaction with both magnetic and electrically sensitive devices at different locations, potentially simultaneously. Finally, the magnetic field focusing between the posts results in extremely strong interactions with any magnetically susceptible material placed there.

The interaction between a single cavity mode and the FMR can be described by two coupled harmonic oscillators, for which the Hamiltonian is read as:
\begin{equation}
    \label{eq:H}
    \hat{H} = \hat{H}_c + \hat{H}_m + \hat{H}_{int},
\end{equation}
where $\hat{H}_c=\hbar\omega\hat{c}^\dag\hat{c}$ represents the photonic mode, $\hat{H}_m=\hbar\omega_m\hat{b}^\dag\hat{b}$ the magnon mode, and $\hat{H}_{int}$ is the Zeeman interaction \citep{USC-magnonphoton-46}, which describes the coupling between the two oscillators for this system. $\hbar$ is the reduced Planck constant, $\omega_{c(m)}/2\pi$ is the cavity (magnon) frequency, and $\hat{c}^\dag$ and $\hat{c}$ ($\hat{b}^\dag$ and $\hat{b}$) are the creation and annihilation cavity (magnon) operators, respectively.

Following \citep{USC-magnonphoton-46}, and as demonstrated in appendix \ref{appendix:PD}, the physics system is described by the Dicke model reading as:
\begin{equation}
    \label{eq:Dicke}
    \hat{H}/\hbar = \omega\hat{c}^\dag\hat{c} + \omega_b\hat{b}^\dag\hat{b} + g(\hat{c}^\dag + \hat{c})(\hat{b}^\dag + \hat{b}).
\end{equation}

An easy way to solve eigenvalues of the Dicke Hamiltonian is to use the Rotating Wave Approximation (RWA) where the counter-rotating terms, $\hat{c}^\dag\hat{b}^\dag$ and $\hat{c}\hat{b}$, are neglected.
In the case of a system being in the USC regime, it is well known \citep{ReviewUSC, ReviewUSC2} that this approximation no longer describes this system. Using the Hopfield-Bogoliubov transformation allows one to solve for the system eigenfrequencies whilst considering co-rotating, $\hat{c}^\dag\hat{b}$ and $\hat{c}\hat{b}^\dag$, and counter-rotating terms:
\begin{equation}
    \label{eq:CMP2}
    \omega_\pm = \dfrac{1}{\sqrt{2}}\sqrt{\omega^2 + \omega_m^2 \pm \sqrt{\left(\omega^2 - \omega_m^2\right)^2 + 16g^2\omega\omega_m}}.
\end{equation}
Moreover, the coupling strength is defined as \cite{Bourhill2020}:
\begin{equation}
    \label{eq:g3}
    \dfrac{g}{2\pi} = \frac{\gamma}{4\pi}\eta\sqrt{\dfrac{\mu_0S\hbar\omega}{V_{m}}}=\eta\sqrt{\omega}\frac{\gamma}{4\pi}\sqrt{\frac{\mu}{g_l\mu_B}\mu_0\hbar n_s},
\end{equation}
where $\gamma=2\pi~28$ GHz.T$^{-1}$ is the gyromagnetic ratio for YIG, $g_l=2$ is the Land\'e g-factor for an electron spin, $\mu_0$ is the vacuum permeability, $\mu_B$ is the Bohr magneton, $\mu=5\mu_B$ is the magnetic moment of the sample, $n_s = 4.22\times10^{27}~\text{m}^{-3}$ is the spin density for YIG \cite{Bourhill2020}, and $\eta$ is the filling factor, where

\begin{equation}
    \label{eq:eta2}
    \eta = \sqrt{\dfrac{\left(\int_{V_m}\mathbf{h}\cdot\hat{x}\mathrm{d}V\right)^2 + \left(\int_{V_m}\mathbf{h}\cdot\hat{y}\mathrm{d}V\right)^2}{V_m\int_{V_c}|\mathbf{h}|^2\mathrm{d}V}}.
\end{equation}

The filling factor describes the proportion of the $\mathbf{h}$-field (x- and y-axis components), perpendicular to the static magnetic field ($\mathbf{H}$-field), named $\mathbf{H}_0$ in \figref{fig:hybrid} 
compared to the $\mathbf{h}$-field for all directions inside the entire cavity volume $V_c$.

\section{Optimization}
\label{sec:Hybrid_system}
An appropriate optimization of the cavity allows one to maximize the coupling and to obtain a quasi-homogeneous $\mathbf{h}$-field inside the YIG slab. With the use of Finite Element Modeling (FEM) and following the procedure described by Bourhill et al. \cite{Bourhill2020}, we were able to precisely predict and therefore optimize prior to construction, the cavity frequency, frequency tuning range, and the coupling strength considering equation (\ref{eq:g3}).

The optimization of the cavity design was based on the maximization of the filling factor $\eta$ and the $\mathbf{h}$-field homogeneity at the first BM inside the YIG slab. For a correct distribution of the RF field inside the cavity (seen as a Perfect Electric Conductor, PEC), it is necessary to consider the electrical property of the YIG, namely a relative dielectric permittivity of 15.  Dynamic magnetic properties are not useful at this stage and instead of considering the magnetic permeability with the Polder tensor, we consider it as that of vacuum.

There exist only three free parameters for the optimization of the hybrid system, two for the size of the posts, the width $W$ and the length $L$, and one for the cavity, the radius $R$. The other parameters such as the height of the cavity and the distance between the posts were fixed by the constraints imposed by the YIG dimension and the cavity manufacturing accuracy. The optimization step is described in appendix \ref{appendix:Opti}, and the optimized values are $W=0.6$ mm, $L=6$ mm, and $R=12$ mm.

\begin{figure}[ht!]
    \centering
    \includegraphics[width=\linewidth]{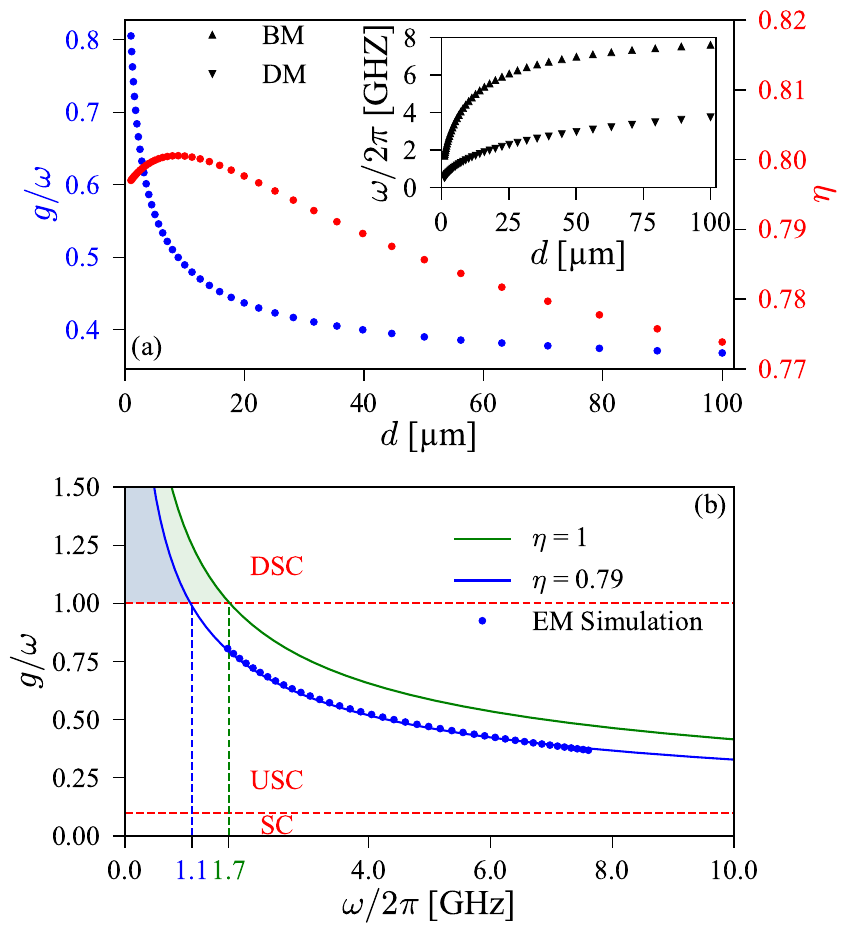}
    \caption{ (a) Evolution of the ratio $g/\omega$ (with $\omega=\omega_{BM}$) in blue and $\eta$ in red versus $d$ for Eigen-Mode simulations (EM). Inset: Evolution of simulated DM and BM frequencies versus $d$. (b) Evolution of $g/\omega$ versus the BM frequency for $\eta=1$ and $\eta=0.79$.}
    \label{fig:optimization}
\end{figure}


The simulated evolution of the two eigenmodes (DM and BM) are shown in the inset of \figref{fig:optimization} (a) with respect to the distance $d$ between posts and the lid of the cavity, with a range from 1 to 100 $\mu$m. Decreasing $d$ will decrease the frequency of the eigenmodes and the frequency difference between the BM and the DM. \figref{fig:optimization} (a) shows electromagnetic simulation results for $\eta$ (right y-axis) and $g/\omega$ (left y-axis) versus $d$ for a cavity with the optimized dimensions, where $\omega=\omega_{BM}$ the frequency mode of interest in our study. $\eta$ is maximized for $d=9~\mu$m. The variation of $\eta$ over this range of $d$ values is only is 2.7\%, therefore we may consider it more or less invariant. The tuneability of the distance $d$ plays a role on the $g/\omega$ ratio as shown in \figref{fig:optimization} (a). Indeed, $\omega/2\pi$ is decreasing with $d$, and $\eta$ is remaining almost constant. Considering Eq. \eqref{eq:g3}, $g/2\pi$ is a function of $\eta$ and the square root of $\omega$. Therefore, the ratio $g/\omega$ will increase with the inverse of the square root of $\omega$ from 36.8 to 80.5\% as $d$ decreases from 100 to 1 $\mu$m.

Fig. \figref{fig:optimization} (b) illustrates the SC to DSC transition for YIG with the frequency dependence of $g/\omega$. The blue dots correspond to the values extracted from EM simulation already discussed in \figref{fig:optimization} (a) and the solid line dependencies are based on equation (\ref{eq:g3}) for two constant values of $\eta$, 0.79 (blue) and 1 (green). The magnetic properties of YIG require working in a specific frequency range in order to explore the DSC. For the maximum reachable value of $\eta$ (green line), which corresponds to the entire $\mathbf{h}$-field perpendicular to $\mathbf{H}_0$ and fully confined to $\mathbf{V}_m$, DSC is possible when the magnons are coupled to a microwave mode below 1.72 GHz \cite{Bourhill2020}. In our case (with $\eta$ close to 0.79), DSC is achievable but at a smaller resonant frequency (1.07 GHz). Note that the optimized cavity configuration of this work does not allow to reach the DSC due to the presence of the dark mode which contaminates the low frequency response and the difficulty to control distance $d$ lower than 3 $\mu$m.

\section{Results and discussion}
\label{sec:R&S}

\begin{figure*}[htb]
    \centering
    \includegraphics[width=\textwidth]{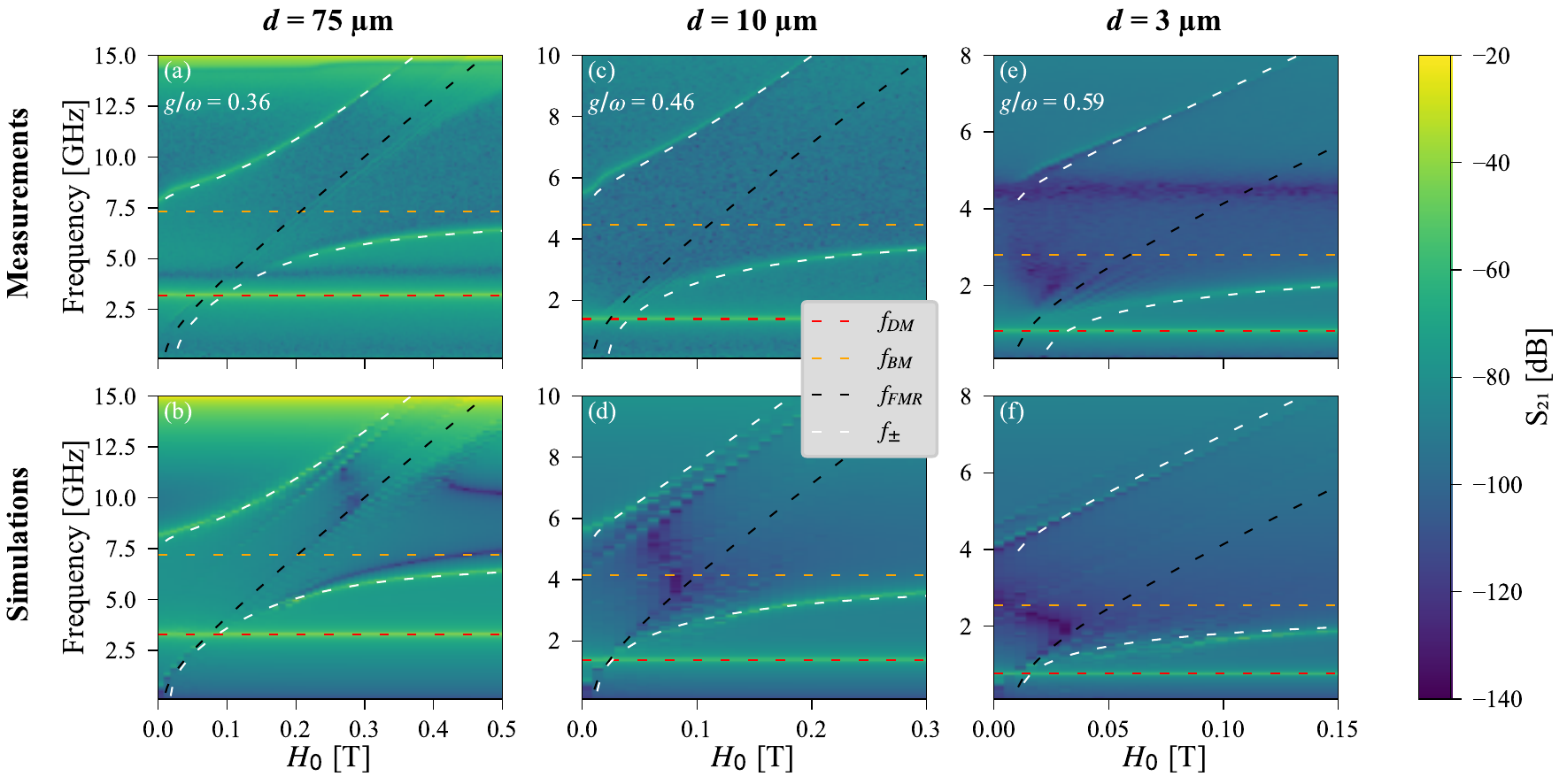}
    \captionsetup{width=\textwidth}
    \caption{ Transmission spectra versus frequency and $H_0$ for (a), (c), (e) measurements at RT, and (b), (d), (f) simulations. Comparison spectra between measurement and simulations are shown for different distances $d$ as labelled. A fit with the Dicke model with a shifted magnon frequency is shown superimposed on (c) and (d) where the FMR frequency ($f_{FMR}=\omega_{FMR}/2\pi$) is shown in black, the DM frequency ($f_{DM}=\omega_{DM}/2\pi$) in red, the BM frequency ($f_{BM}=\omega_{BM}/2\pi$) in orange and the two polariton frequencies ($f_\pm=\omega_{\pm}/2\pi$) in white.}
    \label{fig:FIG2}
\end{figure*}

\subsection{Simulation details}
To compare the experimental results, simulations in the frequency domain (FD), solving for the $S_{21}$ scattering parameter were conducted for different values of $d$ from 2 to 100 $\mu$m. For these simulations, we considered the excitation probes and hence the coupling losses. Losses due to finite conductivity of the cavity walls are also taken into consideration.

The static and dynamic magnetic properties of YIG are used to solve the frequency response of the entire system as a function of the applied magnetic field. The spin dynamics of ferrimagnetic systems can be described by the Landau-Lifshitz-Gilbert (LLG) equation and the frequency dependence of the coupled dynamics can be accurately estimated by using a linear solution of the LLG equation in solving Maxwell’s equations. Some consideration regarding the shape of the YIG sample must be taken into account. The FMR dispersion for a relatively thick slab geometry requires careful consideration. Based on the works of Kittel \citep{Kittel-FMR}, R. I. Joseph and E. Schl\"omann \citep{Joseph-demag}, the demagnetizing field expression has been adapted to our non-ellipsoidal sample of YIG (as described in Appendix B). From these results, it is determined that the demagnetizing field is significantly different from the thin-film form, and therefore for accurate simulations proper consideration of this difference must be taken into consideration. Hence, the effective static magnetic field in the YIG is different from the applied one and read as:
\begin{equation}
    H_i = H_0 - N_{zz}\left(x, y, z\right)M_0
\end{equation}
where $H_i$ is the internal static magnetic field along the z-axis, and $N_{zz}$ is the spatially dependent demagnetizing component along the z-axis, and is describe in Eq. \ref{diag} in appendix \ref{appendix:FMR}.

\subsection{Experimental set-up}

To reach the specifications described above, an aluminum cavity with an accuracy of 20 $\mu$m has been machined.

For the applied static magnetic field, we used an electromagnet where the produced field is aligned along the z-axis (see \figref{fig:hybrid}), in the direction of the height of the posts. $\mathbf{H}_0$ aligns all the spin moments along the z-axis and to saturate the macroscopic YIG magnetization. With the shape of the cavity, the $\mathbf{h}$-field for the BM, considered as the perturbative field, is only along the x-axis inside the YIG slab between the two posts, as shown in \figref{fig:hybrid} (b) due to the constructive interference of the two $\mathbf{h}$-fields around each post.
A gaussmeter allows one to measure in \textit{situ} $H_0$ magnitudes. $S$-parameters are measured with a two-port Vector Network Analyzer (VNA), with the magnitude and phase of the scattering parameters recorded between 0.1 to 15 GHz with an input power of $-10$ dBm.
All measurements are conducted at RT

The magnitude of the $S_{21}$ transmission spectra as a function of $H_0$ are displayed in \figref{fig:FIG2} for measurement and simulation with differing sized gaps between the top of the posts and the roof of the cavity. Experimentally, this is varied by using different cavity lids which had recesses of differing heights machined into them.

\subsection{Results}
\label{sub:Expresults}

    Measurement and simulation results of magnetic spectroscopy of the cavity magnon system are shown in \figref{fig:FIG2} as the first and second row, respectively, for different values of $d$. Each column represents a comparison between a measurement and a simulation with a distance close to the measured value. The latter can be determined by the unperturbed value of $f_{DM}$, which acts as a calibration for $d$.

The external magnetic field was always applied symmetrically for negative and positive values. 
This allows to improve the fit accuracy on measurements, because we have twice as many data points. All measurements with complete frame are shown in appendix \ref{appendix:measurements}.

We can easily distinguish the two hybrid eigenfrequencies $f_+=\omega_+/2\pi$ (for the higher branch) and $f_-=\omega_-/2\pi$ (for the lower branch) from either side of the BM frequency. It should be noted that at low $H_0$ values the BM is not visible, whilst we can clearly see the DM which is the lowest frequency mode and has a negligible coupling with the magnon mode, hence is constant versus $H_0$.

Some minor discrepancies between simulation and experiment should be pointed out:
($i$) an inflection point on the curvature of the upper CMP in frequency at low $H_0$ (observed only in the USC regime) for measurements, appearing neither in simulation nor analytic fits;
($ii$) anti-resonances only appearing either in measurement, the horizontal one around 4.3 GHz in \figref{fig:FIG2} (a) and (e), or in simulation with a $S$-like shape, around 10, 4, and 2 GHz in respectively \figref{fig:FIG2} (b), (d) and (f).
 Let us notice that this anti-resonance does not appear in measurements when a cavity mode is overlapping with this transmission dip, as shown for $d$ = 10 µm in \figref{fig:FIG2} (c) and \figref{fig:CAV01} (e), and for $d$ = 116 µm in \figref{fig:CAV01} (a);
($iii$) another magnon mode exists near the upper CMP in simulations. It is clear that it is another magnon mode because its $\mathbf{H}$-field's frequency dependence does not change as $d$ is varied.\\
Differences given in the two last point could be explained by the fact that the YIG sample is a perfect rectangular prism in simulation whereas the real sample is not. The imperfections of the YIG geometry could result in a weak transmission, which  could be not detected in measurement.

Despite these minor deviations, the agreement between simulations and measurements on the magnon-photon coupling and the resulting CMPs is excellent. In particular, we validated the spatial distribution of the demagnetizing field, hence the expression of the FMR for a slab, and the ability of the Maxwell's equations to describe the system. This permits one to conduct a simulation with a magnetic field larger than experimentally possible in order to extract the BM frequency. Indeed, it is impossible to measure the unperturbed BM frequency $f_{BM}$ in the USC regime even when applying a high magnetic field near to 2 T.

\subsection{Model Description}
\label{sub:ModelDescription}

In the USC regime, the Tavis-Cummings model becomes no longer applicable \citep{BreakJC, USCfirstEXP}, as $g/\omega>0.1$ leads to a failure of the rotating wave approximation as the interaction term of the Hamiltonian can no longer be assumed to be {``}slowly rotating{''} compared to the system terms. The standard model for cavity magnonics is the Dicke model (see Eq. \eqref{eq:Dicke}). However, we have noticed that in the coupling regime of our system, even the Dicke model cannot describe observed polariton frequency dispersion for measurements and simulations, as shown in \figref{fig:Dicke} in appendix \ref{appendix:Dicke}. Another standard model describing light-matter interactions is the Hopfield model \citep{Hopfield1958}, similar to the Dicke model with an additional diamagnetic term. This well known model neither fit the measured data with the use of the Hopfield model, as shown in appendix \ref{appendix:Hopfield}.

To remedy these issues, it has been proposed to modify the Dicke model with the addition of a $\mathbf{H}$-field in the FMR dispersion equation \citep{ApproachDSC}. We also modified the term of the FMR frequency dependence in equation (\ref{eq:CMP2}) to:
\begin{equation}
    \omega_m \rightarrow \omega_m + \Delta_m
\end{equation}
where $\Delta_m=2\pi f_\Delta$ is a frequency shift, which will be further discussed in section \ref{subsection:Discussion}.
This modified Dicke model was found to fit best the experimental and simulation spectra, as seen in the white dash lines of \figref{fig:FIG2} for $d=75$ $\mu$m in (a) and (b), $d=10$ $\mu$m in (c) and (d) and $d=3$ $\mu$m in (e) and (f). Measurement fit, shown in \figref{fig:FIG2} (a), (c), and (e), is achieved with the BM frequency $f_{BM}$ (in orange), the coupling strength $g/2\pi$, and the added frequency $f_\Delta$ as fitting parameters. For simulation fit, shown in \figref{fig:FIG2} (b), (d), and (f), the BM frequency is considered as fixed parameter. Indeed, simulations were performed at an artificial high $\mathbf{H}$-field ($H_0=10$ T), in order to tune the magnon mode many orders of coupling strength away, and clearly distinguish the two photonic modes.

An offset far detuned from the BM frequency at a zero $\mathbf{H}$-field arises for high $g/\omega$ when the FMR is shifted. When the FMR is not shifted, in the standard Dicke model, no BM frequency detuning at zero $\mathbf{H}$-field exists for any $g/\omega$ value. See appendix \ref{appendix:gap} for more details about the frequency detuning at zero $\mathbf{H}$-field.

All values of fit parameters, for measurements and simulations, are available in appendix \ref{appendix:measurements}, and are pooled in \figref{fig:FIG3}. For the measurements (shown in blue), the distance $d$ has been estimated from the measured DM frequency. The fitted BM frequencies of the measurements are in good agreement with simulations (shown in black in the inset of \figref{fig:FIG3}). Regarding the coupling strength $g/\omega$, we achieve a ratio $g/\omega$ ranging from 0.35 to 0.59, corresponding to $d=116$ $\mu$m to $d=4$ $\mu$m, respectively. As mentioned in section \ref{sec:Hybrid_system}, the values of $g/\omega$ are different from the optimization step ones (dotted red curve), mainly due to the different estimated frequencies, shown in the inset. Once again, the correlation between fitted simulations and measurements for the ratio $g/\omega$ are also good. This clearly demonstrates the validity of the simulations.

\begin{figure}[h!]
    \centering
    \includegraphics[width=\linewidth]{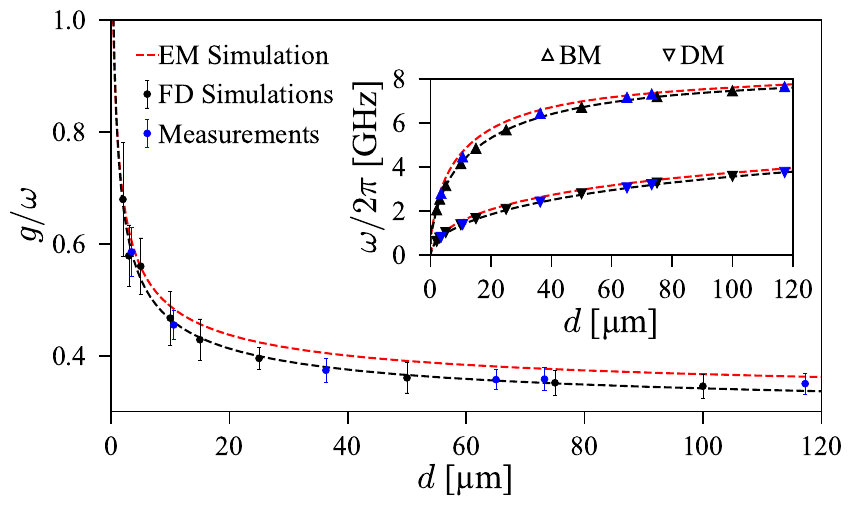}
    \captionsetup{width=\linewidth}
    \caption{ $g/\omega$ versus $d$ for fitted FD simulations (in black) and for fitted measurements (in blue). The simulation trend is plotted in the black, dashed line. Inset: DM and BM frequencies versus the distance $d$, in the black dashed line are shown DM and BM reading values from simulations at extremely high applied $\mathbf{H}$-field. Eigen-Mode (EM) simulations are shown as the red dashed line.}
    \label{fig:FIG3}
\end{figure}

\subsection{Discussion}
\label{subsection:Discussion}

We discuss here the physical meaning of the frequency shift in the modified Dicke model. For a deeper understanding of the behavior of this added term, we investigated the transition between the SC and the USC regimes. In order to study a wide range or $g/\omega$ values, we have used two other cavities with the same YIG sample.
The first described machined cavity will be named {``}CAV$_{01}${''} in the following. This cavity operates in a $g/\omega$ range from 0.35 to 0.59, as mentioned in the table \ref{tab:CAV01} in appendix \ref{appendix:measurements}.

The second cavity, {``}CAV$_{02}${''}, has been 3D printed and has the same shape as CAV$_{01}$, but with smaller height posts. This cavity is performing in a certain range of $g/\omega$, from 0.28 to 0.32 (see table \ref{tab:CAV02} in appendix \ref{appendix:measurements}).

The third cavity, {``}CAV$_{03}${'"}, is also a double re-entrant 3D printed cavity with cylinder posts, adjustable in height. This cavity was used in a previous work \citep{Bourhill2020} to experimentally verify a reworked theory that predicts coupling values from simulations alone. The cavity has radius $R_{cav}$ = 20 mm and height $H_{cav}$ = 4.6 mm, whilst the posts have radius $R_{post}$ = 2.05 mm and are spaced to 2.7 mm. The operating ratio $g/\omega$ is lower than the two others cavities and enables to have experimental results at the SC/USC threshold, with $g/\omega$ comprised between 0.12 and 0.25 (see table \ref{tab:CAV03} in appendix \ref{appendix:measurements}).

The operating range in BM frequencies, coupling strengths, and added frequencies for the three cavities are summarized in Table \ref{tab:table1}.
\begin{table}[h!]
    \caption{\label{tab:table1}%
        Operating range of the cavities}
    \begin{ruledtabular}
        \begin{tabular}{cccc}
            \textrm{Cavity}         &
            \textrm{$f_{BM}$ [GHz]} &
            \textrm{$g/2\pi$ [GHz]}      &
            \textrm{$\Delta_m/2\pi$ [GHz]}                                         \\
            \colrule
            \textrm{CAV$_{01}$}     & 2.80 - 7.65 & 1.64 - 2.68   & 2.27 - 2.59 \\
            \textrm{CAV$_{02}$}     & 7.63 - 9.79 & 2.42 - 2.72 & 1.63 - 1.74 \\
            \textrm{CAV$_{03}$}     & 2.35 - 5.53  & 0.58 - 0.69  & 0.29 - 0.50   \\
        \end{tabular}
    \end{ruledtabular}
\end{table}

Thanks to the validation of the FD simulations, we were able to simulate the USC CAV$_{01}$ design for different dimensions of the YIG slab, while keeping the aspect ratio of the slab constant. Since the demagnetizing components described in Eq. \eqref{diag} are only dependent on this aspect ratio, the FMR remains unchanged.
However, still decreasing the YIG slab dimensions decreases the filling factor $\eta$, therefore the coupling strength and $g/\omega$ from 63 to 5 $\%$ with $d=50$ $\mu$m.

\begin{figure}[h!]
    \centering
    \includegraphics[width=\linewidth]{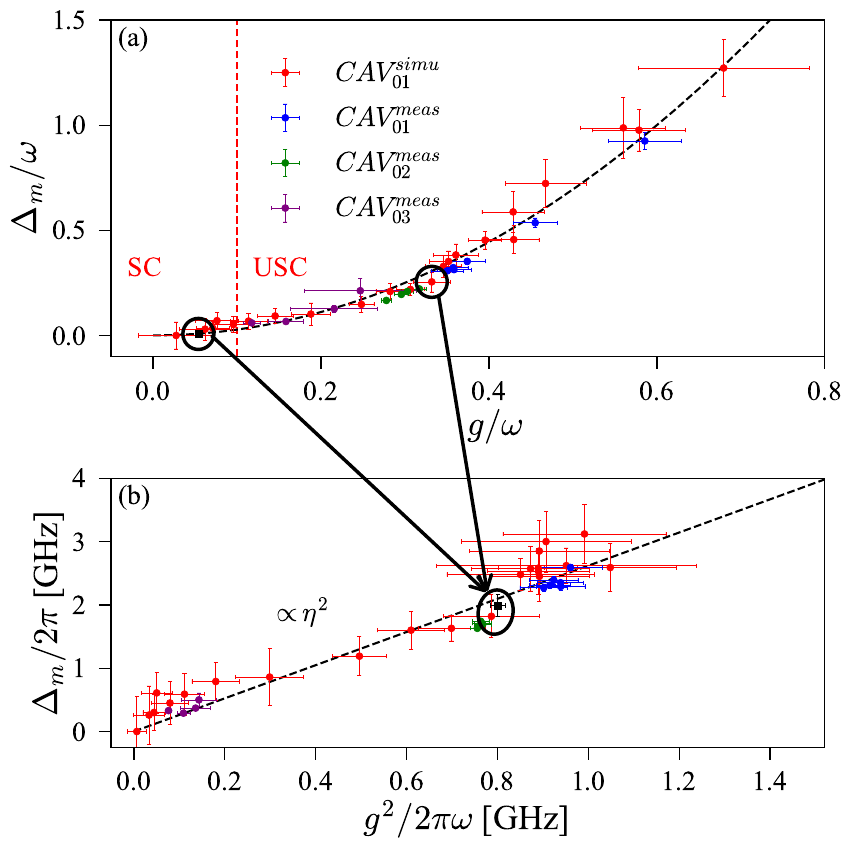}
    \captionsetup{width=\linewidth}
    \caption{(a) $\Delta_m/\omega$ versus $g/\omega$ and (b) $\Delta_m/2\pi$ versus $g^2/2\pi$.  Shown are the FD simulations on CAV$_{01}$ in red and measurements in blue, in green for CAV$_{02}$, and in purple for CAV$_{03}$. In (a) and (b), fitted values for a reduced CAV$_{01}$ with an aspect ratio equal to 0.025 as the black square.
    The two data points circled in (a) corresponds to the same value of $\Delta_m/2\pi$ in (b).}
    
    \label{fig:FIG4}
\end{figure}

We plotted $\Delta_m/\omega$ versus $g/\omega$ in \figref{fig:FIG4} (a) which clearly display a quadratic dependence. For $g/\omega \leq 0.1$, $\Delta_m/\omega$ is more or less negligible. This description agrees with the commonly situated transition point (shown as the red dotted line) between the SC and USC regimes where all models converge. Our simulations show the need for the $\Delta_m/2\pi$ parameter to properly fit the data. \figref{fig:FIG4} (b) shows $\Delta_m/2\pi$ versus $g^2/2\pi\omega$ which is proportional to the square of the filling factor, $\eta^2$. 
According to this observation and the definition of $\eta$, we noticed that the more this energy is confined in the YIG, the larger the shift in the magnon frequency will be. In the literature, the parameter $\eta$ is not so often considered or estimated. In ref \citep{Bourhill2020}, we had the opportunity to test the model of equations (\ref{eq:g3}) and (\ref{eq:eta2}) on multiple published experimental results, and $\eta$ rarely exceeds 0.05 in any of them. As a reminder, and in view of the description in \figref{fig:optimization}, our system (CAV$_{01}$) proposes a $\eta$ of about 0.79.

In \figref{fig:FIG4} is represented by square marker a simulation where dimensions of the cavity and the YIG are reduced by a ratio equal to 0.025 for $d=50$ $\mu$m. By decreasing the dimensions of the entire cavity CAV$_{01}$ by this ratio, the BM frequency is increased to 275 GHz. Then, this cavity operates in the SC regime. However, the proportion of the $\mathbf{h}$-field in the YIG remains the same, hence also $\eta$.

In (a), are circled the reduced system performing in the SC regimes, and the unmodified cavity in the USC regime presenting the same $\eta$ value.
In (b), it is shown that the frequency shift $\Delta_m/2\pi$ is the same for both cavities, and for the same value of $g^2/2\pi\omega$. Considering eq. \ref{eq:g3}, $\Delta_m/2\pi$ depends of the magnetic properties of the YIG and $\eta^2$. It is then important to note that this effect is not bounded to the coupling strength and hence to the coupling regime, but instead to the filling factor, something that has never been discussed so far.

As physical mechanism of $\Delta_m/2\pi$, the nonlinear optical processes having similar behavior, would be a good candidate. Among them, two different effects attracted our attention: the multi-photon Rabi oscillations \citep{ReviewUSC, PhysRevA.95.063849, Garziano2015} for its effective coupling being proportional to $g^2/\omega$. When the coupling between an artificial single atom and a cavity is in the USC regime, the system can exchange several photons (and undergo multi-photon Rabi oscillations) instead to a single one (commonly known as Rabi oscillation); and the self-Kerr, and the cross-Kerr effects \citep{Yang2021, Bi2021, Yang2022, Wu2021, Zhang2019} presenting a frequency shift of the magnon, due to magnetocrystalline anisotropy and magnon-magnon interactions, respectively.

\section{Conclusion}

In conclusion, we proposed a double re-entrant cavity design to achieve USC magnon/photon coupling at microwave frequencies, which was supported by both experimental data and electromagnetic simulations. To the best of our knowledge, this is the only demonstration of USC magnon/photon coupling at room temperature so far. Noteworthily, reaching the USC without cryogenic temperatures is promising for the development of RF applications based on cavity spintronics.

We explained the importance of optimizing the filling-factor $\eta$ for reaching the USC, aside from just the frequency of the resonator and the spin density. Importantly, the cavity we proposed is parametrized by the distance $d$ between the posts and the lid. We showed that tuning this parameter allowed to continuously go from the regular SC to the USC regime.
The ability to study the transition from the SC to USC regime is a significant step towards understanding the physics of USC magnon/photon coupling. 

Indeed, we showed that the standard models describing the coupling of a single resonator mode to many dipoles (e.g. the Dicke and Hopfield models) failed to properly decsribe our experimental data. Nevertheless, thanks to the validation of our electromagnetic simulations, we showed that a frequency shift in the magnon frequency adequately modelled our data, which we note is fully captured by the classical Maxwell's equations. Furthermore, we showed that this frequency shift only depended on the filling-factor $\eta$, highlighting its importance for hybrid magnon/photon systems. While the physical origin of the magnon's frequency shift is still unknown, we hope that its relation with $\eta$ will motivate further research into deriving a proper theoretical model for USC magnon/photon coupling.

\section*{Acknowledgments}
This work is part of the research program supported by the European Union through the European Regional Development Fund (ERDF), by Ministry of Higher Education and Research, Brittany and Rennes M\'etropole, through the CPER Project SOPHIE/STIC $\&$ Ondes, by the CPER SpaceTechDroneTech, by Brest M\'etropole, and the ANR project MagFunc. JB is funded by the Australian Research Council Centre of Excellence for Engineered Quantum Systems, CE170100009 and the Centre of Excellence for Dark Matter Particle Physics, CE200100008. We thank Alan Gardin for useful discussions.

\bibliography{QAVITY_P01}
\bibliographystyle{unsrt}

\onecolumngrid
\newpage
\appendix

\section{Physics Description}
\label{appendix:PD}

The system under consideration is best described by the Hamiltonian of two coupled harmonic oscillators. The oscillators represent the cavity photonic mode $\hat{H}_c=\hbar\omega\hat{c}^\dag\hat{c}$, and the uniformly precessing Kittel magnon mode $\hat{H}_m=\hbar\omega_m\hat{b}^\dag\hat{b}$, where $\hbar$ is the reduced Planck constant, $\omega/2\pi$ and $\omega_m/2\pi$ are respectively the cavity and magnon frequencies, and $\hat{c}^\dag$ ($\hat{b}^\dag$) and $\hat{c}$ ($\hat{b}$) are the creation and annihilation cavity (magnon) operators. The coupling is then read as a an interaction $\hat{H}_{int}$, and the entire system Hamiltonian can then be written as:
\begin{equation}
    \label{eq:H}
    \hat{H} = \hat{H}_c + \hat{H}_m + \hat{H}_{int}
\end{equation}

The quantization of the Maxwell's equation leads to the expression of the vector potential:
\begin{equation}
    \hat{\mathbf{A}}(\mathbf{r}, t) = \sum_{n}\dfrac{\hat{q}_n(t)}{\sqrt{\epsilon_0\epsilon_{r, n}}}\mathbf{U}_n(\mathbf{r})
\end{equation}
where $\epsilon_0$ the vacuum permittivity, $\epsilon_{r,n}$ the relative permittivity experienced by the cavity mode $n$, $\hat{q}_n(t)$ is the temporal term, and $\mathbf{U}_n(\mathbf{r})$ is the space dependent operator.
This expression is generalized for all modes in a cavity. In the following, we will concern ourselves with only a single mode.

This potential vector in a cavity is comparable to a simple harmonic oscillator, where radiation modes are defined according to annihilation and creation operators:
\begin{equation}
    \begin{aligned}
         & \hat{c} = \dfrac{1}{\sqrt{2\hbar\omega}}\left(\omega\hat{q} - i\hat{\dot{q}}\right)      \\
         & \hat{c}^\dag = \dfrac{1}{\sqrt{2\hbar\omega}}\left(\omega\hat{q} + i\hat{\dot{q}}\right)
    \end{aligned}
\end{equation}

Then, the RF magnetic field ($\mathbf{h}$-field) bounded to the cavity mode is\citep{Rana}:
\begin{equation}
    \label{eq:Hc}
    \hat{\mathbf{h}} = \dfrac{1}{\mu_0}\nabla\times\hat{\mathbf{A}} =\dfrac{1}{\mu_0}\sqrt{\dfrac{\hbar}{2\omega\epsilon_0\epsilon_{r,c}}}\left(\hat{c}^\dag + \hat{c}\right)\nabla\mathbf{\times} \mathbf{U}
\end{equation}
where $\mu_0$ is the vacuum permeability, $\epsilon_0$ the vacuum permittivity, $\epsilon_{r,c}$ the relative permittivity experienced by the cavity mode, and $\mathbf{U}$ is the space dependent operator of the potential vector. The component of this field perpendicular to the sample{'}s magnetization direction will couple to the Kittel magnon mode.

For such a uniform precession of the magnetic sample, we introduce the macrospin operator considering the entire sample, as:
\begin{equation}
    \label{eq:S}
    \hat{\mathbf{S}}=\frac{V_m}{\gamma}\hat{\mathbf{M}}
\end{equation}
where $\hat{\mathbf{M}}$ is the magnetization operator, $V_m$ the volume of the magnetic sample, and $\gamma$ the gyromagnetic ratio.

We consider a saturated magnetization by the use of an applied static magnetic field ($\mathbf{H}$-field) in the z-axis direction. It is then useful to introduce spin raising $\hat{S}^+$ and lowering $\hat{S}^-$ operators. Following the Holstein-Primakoff transformation\citep{Springer2009} and considering low excitation numbers versus the total spin number of the macrospin operator, we obtain:
\begin{equation}
    \label{eq:S2}
    \begin{aligned}
         & \hat{S}^+ = \hat{S}_x + i\hat{S}_y = \sqrt{2S - \hat{b}^\dag\hat{b}}\hat{b} \approx \sqrt{2S}\hat{b}           \\
         & \hat{S}^- = \hat{S}_x - i\hat{S}_y = \hat{b}^\dag\sqrt{2S - \hat{b}^\dag\hat{b}} \approx \sqrt{2S}\hat{b}^\dag \\
         & S_z = S - \hat{b}^\dag\hat{b}
    \end{aligned}
\end{equation}
where $S=\frac{\mu}{g_l\mu_B}N_s$ is the total spin number of the macrospin, $\mu_B$ is the Bohr magneton, $\mu$ is the magnetic moment of the sample, $g_l$ is the Landé g-factor, and $N_s=n_sV_m$ is the number of spins in the sample, with $n_s$ the spin density.

The interaction term corresponds in this case to the Zeeman energy:
\begin{equation}
    \label{eq:Hint}
    \hat{H}_{int} = - \mu_0\int_{V_m}\hat{\mathbf{M}}\cdot\hat{\mathbf{h}}\mathrm{d}V
\end{equation}

Substituting $\hat{\mathbf{h}}$ and $\hat{\mathbf{M}}$ in equation (\ref{eq:Hint}) by their expressions in equations (\ref{eq:Hc}) and (\ref{eq:S}), and replacing cartesian macrospin values by raising and lowering ones, with neglecting z-axis terms, we arrive at:
\begin{equation}
    \label{eq:Hint2}
    \hat{H}_{int}/\hbar = g_x(\hat{c} + \hat{c}^\dag)(\hat{b} + \hat{b}^\dag) + ig_y(\hat{c} + \hat{c}^\dag)(\hat{b} - \hat{b}^\dag)
\end{equation}
where the coupling strengths are defined as:
\begin{equation}
    \label{eq:g}
    \begin{aligned}
         & g_x = - \dfrac{\gamma}{2V_m}\sqrt{\dfrac{\hbar S}{\omega\epsilon_{r,c}\epsilon_0}}\int_{V_m}(\nabla\times\mathbf{U})\cdot\hat{x}\mathrm{d}V \\
         & g_y = \dfrac{\gamma}{2V_m}\sqrt{\dfrac{\hbar S}{\omega\epsilon_{r,c}\epsilon_0}}\int_{V_m}(\nabla\times\mathbf{U})\cdot\hat{y}\mathrm{d}V
    \end{aligned}
\end{equation}

In order to consider the integration of the $\nabla\times\mathbf{U}$ term in the magnetic sample volume in x- and y-axis, it is needed to rewrite it considering the entire $\mathbf{h}$-field {``}seen{''} by the sample. Therefore, it is convenient to normalize the $\mathbf{h}$-field against that of the entire cavity. The classical expression for the $\mathbf{h}$-field energy for a single cavity mode is:
\begin{equation}
    \mathcal{E} = \dfrac{\mu_0}{2}\int\mathbf{h}\cdot\mathbf{h}\mathrm{d}V = \dfrac{\mu_0}{2\epsilon_0\epsilon_{r,c}}q\cdot q\int\left(\nabla\times\mathbf{U}\right)\cdot\left(\nabla\times\mathbf{U}\right)\mathrm{d}V
\end{equation}
where $\int_{V_c}\left(\nabla\times\mathbf{U}\right)\cdot\left(\nabla\times\mathbf{U}\right)=\dfrac{\epsilon_{r,c}\omega^2}{c^2}$\citep{Rana}.

Regarding the ratio of the $\mathbf{h}$-field energy in the magnetic sample versus the one in the whole cavity, we get:
\begin{equation}
    \dfrac{\mathcal{E}_{V_m}}{\mathcal{E}_{V_c}} = \dfrac{\int_{V_m}\mathbf{h}\cdot\mathbf{h}\mathrm{d}V}{\int_{V_c}|\mathbf{h}|^2\mathrm{d}V} = \dfrac{\int_{V_m}\left(\nabla\times\mathbf{U}\right)\cdot\left(\nabla\times\mathbf{U}\right)\mathrm{d}V}{\epsilon_{r,c}\omega^2/c^2}
\end{equation}

Using the center and right terms of the above equation, deriving numerators and applying the square root, we finally read the infinitesimal normalized energy amplitude of the $\mathbf{h}$-field:
\begin{equation}
    \label{eq:rotU}
    \nabla\times\mathbf{U}=\dfrac{\sqrt{\epsilon_{r,c}}\omega}{c}\dfrac{\mathbf{h}}{\sqrt{\int_{V_c}|\mathbf{h}|^2\mathrm{d}V}}
\end{equation}

Equation (\ref{eq:H}), with the use of equation (\ref{eq:Hint2}), can be rewritten over a matrix form as:
\begin{equation}
    \begin{aligned}
        \hat{H} & = \dfrac{1}{2}
        \begin{bmatrix}
            \hat{c}^\dag & \hat{b}^\dag & \hat{c} & \hat{b}
        \end{bmatrix}
        H
        \begin{bmatrix}
            \hat{c}^\dag & \hat{b}^\dag & \hat{c} & \hat{b}
        \end{bmatrix}^\dag
        + const                  \\
        H       & =
        \begin{bmatrix}
            \omega  & g_x + ig_y & 0         & g_x - ig_y \\
            g_x - ig_y & \omega_m  & g_x - ig_y & 0         \\
            0         & g_x + ig_y & \omega  & g_x - ig_y \\
            g_x + ig_y & 0         & g_x + ig_y & \omega_m
        \end{bmatrix}
    \end{aligned}
\end{equation}

Using Hopfield-Bogolubov transformation \citep{ApproachDSC}, the solution of the problem is to find polariton operators $\hat{p}_\pm$, expressed as a linear combination of $\hat{c}$, $\hat{c}^\dag$, $\hat{b}$, and $\hat{b}^\dag$. Being bosonic operators, they should obey the Hopfield formulation \citep{Hopfield1958}:
\begin{equation}
    \label{eq:comutator}
    \left[\hat{p}_\pm, \hat{H}\right] = \omega_\pm\hat{p}_\pm
\end{equation}
where $\omega_\pm/2\pi$ are frequency eigenvalues associated with the eigen-operators $\hat{p_\pm}$.

As previously, the Hamiltonian in the polaritonic basis can be rewritten as:
\begin{equation}
    \hat{H} = \dfrac{1}{2}
    \begin{bmatrix}
        \hat{p}_-^\dag & \hat{p}_+^\dag & \hat{p}_- & \hat{p}_+
    \end{bmatrix}
    M
    \begin{bmatrix}
        \hat{p}_-^\dag & \hat{p}_+^\dag & \hat{p}_- & \hat{p}_+
    \end{bmatrix}^\dag
\end{equation}
In order to respect equation \ref{eq:comutator}, the Hopfield matrix $M$ has to be read as:
\begin{equation}
    M = H\mathrm{diag}\left(1, 1, -1, -1\right)
\end{equation}

Solving eigenvalues of the matrix $M$ leads to:
\begin{equation}
    \label{eq:CMP}
    \omega_\pm = \dfrac{1}{\sqrt{2}}\sqrt{\omega^2 + \omega_m^2 \pm \sqrt{\left(\omega^2 - \omega_m^2\right)^2 + 16g^2\omega\omega_m}}
\end{equation}
where the coupling strength $g = \sqrt{g_x^2 + g_y^2}$ is defined as:
\begin{equation}
    \label{eq:g2}
    \begin{aligned}
        \dfrac{g}{2\pi} = \frac{\gamma}{4\pi}\eta\sqrt{\dfrac{\mu_0S\hbar\omega}{V_{m}}}=\eta\sqrt{\omega}\frac{\gamma}{4\pi}\sqrt{\frac{\mu}{g_l\mu_B}\mu_0\hbar n_s}
    \end{aligned}
\end{equation}
with the filling factor:
\begin{equation}
    \eta = \sqrt{\dfrac{\left(\int_{V_m}\mathbf{h}\cdot\hat{x}\mathrm{d}V\right)^2 + \left(\int_{V_m}\mathbf{h}\cdot\hat{y}\mathrm{d}V\right)^2}{V_m\int_{V_c}|\mathbf{h}|^2\mathrm{d}V}}
\end{equation}

It is important to note that eigenfrequencies are solution of the Dicke model\citep{ReviewUSC, ReviewUSC2}. Finally the Hamiltonian can be rewritten over the easier form:
\begin{equation}
    \hat{H}/\hbar = \omega\hat{c}^\dag\hat{c} + \omega_b\hat{b}^\dag\hat{b} + g(\hat{c}^\dag + \hat{c})(\hat{b}^\dag + \hat{b})
\end{equation}

\section{Cavity Optimization}
\label{appendix:Opti}

\begin{figure}[ht!]
    \centering
    \includegraphics[width=0.482352941\textwidth]{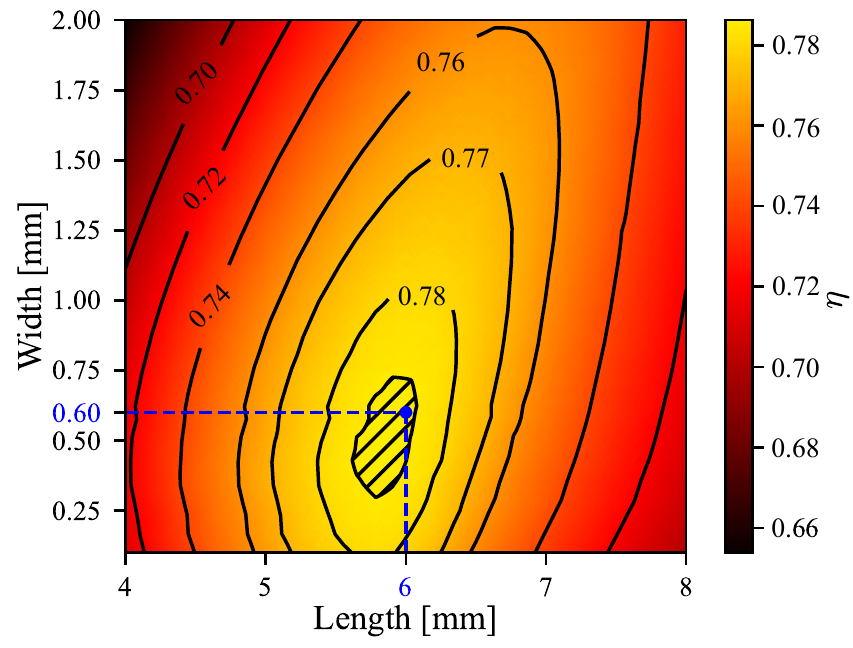}
    \captionsetup{margin={0.\textwidth, 0.517647058823\textwidth}}
    \caption{ filling factor $\eta$ function of the width ($W$) and the length ($L$) of the two posts}
    \label{fig:optieta}
\end{figure}

\figref{fig:optieta} is a representation of the optimization of the filling factor $\eta$ for two of the variable parameters; the width ($W$) and the length ($L$) of the posts, with $d$ chosen equal to 50 µm. The cavity radius ($R$) has been chosen at its optimized value. The containment of the $\mathbf{h}-field$ inside the YIG is at its maximum when the post dimensions are of the slab dimensions. Hence, the width of the posts has been optimized over a range from 0.1 mm to 2 mm, and their lengths from 4 mm to 8 mm. The radius of the cavity does not have a big impact on $\eta$. The cavity radius has been optimize over a range from 10 mm to 14 mm.

Each contour represents the value of $\eta$ with respect to $W$ and $L$. The hashed contour delimits the surface where $\eta\ge78.5\%$. For better feasibility, we choose the largest values of $W$ and $L$. This leads to an optimal value of $\eta$ for $W=0.6$ mm, $L=6$ mm, and $R=12$ mm.

\section{FMR model}
\label{appendix:FMR}
Using the Landau-Lifshitz equation of the magnetization with the proper approximations leads us to the FMR pulsation for all types of ferromagnet shapes \citep{Kittel-FMR}:

\begin{equation}
    \omega_0 = \gamma \sqrt{\left(\dfrac{\omega_e}{\gamma}\right)^2 - \left[\left(N_{xy} + N_{yx}\right) M_s\right]^2}\label{FMR}
\end{equation}
where $M_s$ is the saturation magnetization, $N_{i, j}$ is a component of the demagnetizing tensor at  the $i^{th}$ column and the $j^{th}$ row, and $\omega_e$ is the FMR pulsation for an ellipsoidal body read as:
\begin{equation}
    \omega_e = \gamma \sqrt{\left[|H_z| + \left(N_{xx} - N_{zz}\right) M_s\right] \left[|H_z| + \left(N_{yy} - N_{zz}\right) M_s\right]}\label{FMRellipsoid}
\end{equation}\\
Using the perturbation theory with a small perturbation on the $\mathbf{H}$-field $\epsilon = \dfrac{M_z}{H_z}$ at the first order, it is shown that the demagnetizing components for a rectangular prism are for the diagonal components\citep{Joseph-demag}:
\begin{equation}
    N_{kk}^{(1)} = \dfrac{1}{4\pi} \begin{Bmatrix}
        cot^{-1}\left[f\left(x_i, x_j, x_k\right)\right] + cot^{-1}\left[f\left(-x_i, x_j, x_k\right)\right] +    \\
        cot^{-1}\left[f\left(x_i, -x_j, x_k\right)\right] + cot^{-1}\left[f\left(x_i, x_j, -x_k\right)\right] +   \\
        cot^{-1}\left[f\left(-x_i, -x_j, x_k\right)\right] + cot^{-1}\left[f\left(-x_i, x_j, -x_k\right)\right] + \\
        cot^{-1}\left[f\left(x_i, -x_j, -x_k\right)\right] + cot^{-1}\left[f\left(-x_i, -x_j, -x_k\right)\right]
    \end{Bmatrix}\label{diag}
\end{equation}
with :
\begin{equation}
    f\left(x_i, x_j, x_k\right) = \dfrac{\sqrt{\left(a_i - x_i\right)^2 + \left(a_j - x_j\right)^2 + \left(a_k - x_k\right)^2} \left(a_k - x_k\right)}{\left(a_i - x_i\right) \left(a_j - x_j\right)}\label{ffunc}
\end{equation}\\
and for the off-diagonal terms:
\begin{equation}
    N_{ik}^{(1)} = -\dfrac{1}{4\pi} log
    \begin{Bmatrix}
        \dfrac{G\left(\bm{r}|a_i, a_j, a_k\right) G\left(\bm{r}|-a_i, -a_j, a_k\right) G\left(\bm{r}|-a_i, a_j, -a_k\right) G\left(\bm{r}|a_i, -a_j, -a_k\right)}{G\left(\bm{r}|-a_i, a_j, a_k\right) G\left(\bm{r}|a_i, -a_j, a_k\right) G\left(\bm{r}|a_i, a_j, -a_k\right) G\left(\bm{r}|-a_i, -a_j, -a_k\right)}
    \end{Bmatrix}
    \label{offdiag}
\end{equation}\\
with:\begin{equation}
    G\left(\bm{r}|a_i, a_j, a_k\right) = \left(a_j - x_j\right) + \sqrt{\left(a_i - x_i\right)^2 + \left(a_j- x_j\right)^2 + \left(a_k - x_k\right)^2}\label{gfunc}
\end{equation}

Let us notice that the demagnetizing components are spatially dependent and were averaged to x, y, and z equal to zero for analytical equations. For the YIG dimensions mentioned in the manuscript, the off-diagonal components of the demagnetizing tensor are equal to zero, then the slab as he same FMR frequency as read in Eq. \ref{FMRellipsoid}.

\section{Dicke model}
\label{appendix:Dicke}

\subsection{Normal phase}
The Dicke model is the simplest model to describe the magnon-photon interaction. It consider each Hamiltonian of the cavity photonic mode and magnon as well as the interaction Hamiltonian\citep{ApproachDSC}:
\begin{equation}
    \hat{H} = \omega\hat{c}^\dag\hat{c} + \omega_m\hat{b}^\dag\hat{b} + g(\hat{c}^\dag + \hat{c})(\hat{b}^\dag + \hat{b})
\end{equation}
From this equation, we can easily solve for the eigenmodes, that is to say the polaritronic modes described in Eq. \eqref{eq:CMP}.
This equation is only valid when the ratio $g/\omega$ is less than 0.5. For a description of a system with a ratio higher than 0.5, it is necessary to use the Dicke superradiant phase \citep{Hepp1973}.

\subsection{Superradiant phase}
The superradiant phase is a quantum transition in the Dicke model and represents the displacement of bosonic modes \citep{ApproachDSC}: $\hat{c}^\dag \rightarrow \hat{a}^\dag + \sqrt{\alpha}$ and $\hat{b}^\dag \rightarrow \hat{d}^\dag - \sqrt{\beta}$, where $\alpha$ and $\beta$ represent averaged values of the displaced ground states for the photon and the magnon, respectively. Using Holstein-Primakoff transformation in the Dicke Hamiltonian, the eigen-frequencies become:
\begin{equation}
    \omega_\pm = \dfrac{1}{\sqrt{2}}\sqrt{\omega^2 + \tilde{g}^4\omega_m^2 \pm \sqrt{\left(\omega^2 - \tilde{g}^4\omega_m^2\right)^2 + 4\omega^2\omega_m^2}}
\end{equation}
where $\tilde{g} = 2\frac{g}{\omega}$.\\

\begin{figure}[h!]
    \includegraphics{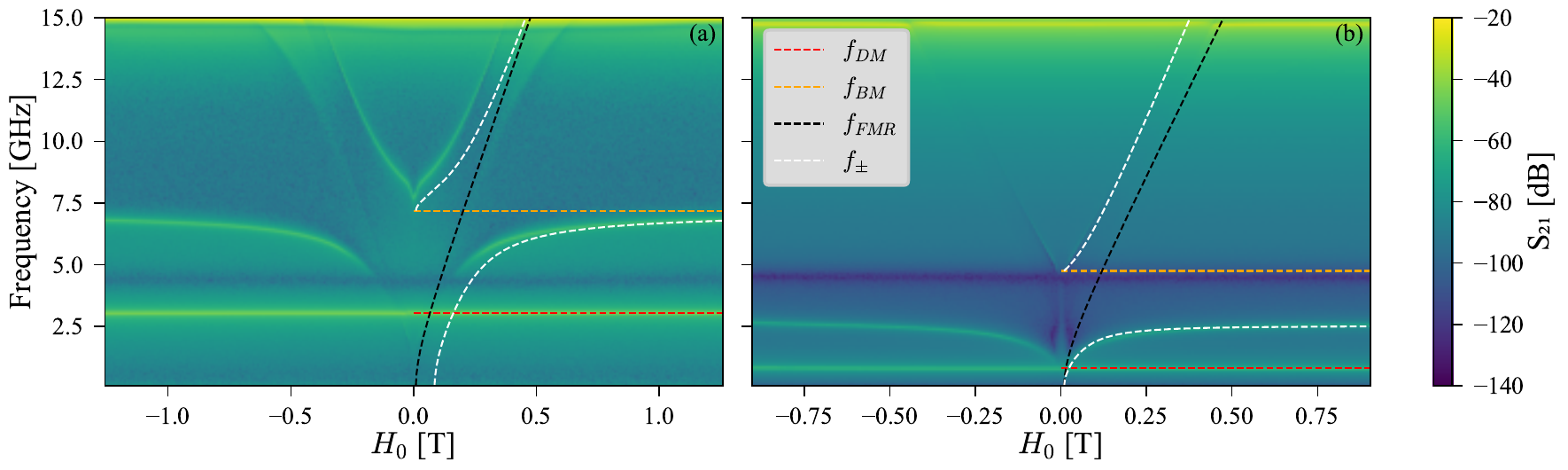}
    \caption{ Transmission spectra versus the RF frequency and the $\mathbf{\mathbf{H}}$-field. Fitted polariton branches are shown in white. The BM frequency (in orange) and the coupling strength are variables. The FMR is shown in black and the DM in red. Fits were be done with the normal phase of the Dicke model. In (a) the measurement for $d=50$ $\mu$m and a fit with the Dicke model. In (b) the measurement for $d=4$ $\mu$m and a fit with the Dicke superradiant phase model.}
    \label{fig:Dicke}
\end{figure}

In this case, \figref{fig:Dicke} (b) shows fitted measurement with the superradiant Dicke model. For this fit, we do not need to add a frequency term on the FMR and we found that $\omega = 4.75$ GHz and $g = 2.58$ GHz.

With comparing BM frequencies, DM frequencies, and $g$ as done in section \ref{sub:Expresults}, the BM frequency and $g$ should be higher than those obtained from simulation. Because of this mismatch, it seems that the superradiant phase is not reached.

\figref{fig:Dicke} shows transmission spectra with respect to the frequency and the $H$-field. A fit has be done with the standard Dicke model. The two eigen-modes of the fit shown in dotted white line which are not consistent with the measurement prove the inability to fit with the standard Dicke model.

\section{Hopfield model}
\label{appendix:Hopfield}

\subsection{Standard}
The Hopfield model is equivalent to the Dicke one with a supplementary term: the diamagnetic one.\\
Considering a carried particle in a magnetic field, we redefine the impulse of the system\citep{ApproachDSC}:
\begin{equation}
    \hat{p} \rightarrow \hat{p} - q\hat{A}
\end{equation}
with $\hat{A}$ the vector potential and associated to the photonic mode: $\hat{A} \propto (\hat{c}^\dag + \hat{c})$\\
We finally have the Hopfield Hamiltonian of the system:
\begin{equation}
    \label{eq:Hopfield}
    \hat{H} = \omega\hat{c}^\dag\hat{c} + \omega_m\hat{b}^\dag\hat{b} + g(\hat{c}^\dag + \hat{c})(\hat{b}^\dag + \hat{b}) + D(\hat{c}^\dag + \hat{c})^2
\end{equation}
where $D$ is the diamagnetic term where the Thomas-Reiche-Kuhn sum rules gives $D=\frac{g^2}{\omega}$.\\
With using Hopfield-Bogolubav transformation and redefining $g$ as $g\sqrt{\frac{\omega_m}{\omega}}$ we have:

\begin{equation}
    \omega_\pm = \dfrac{1}{\sqrt{2}} \sqrt{\omega^2 + \omega_m^2 + 4g^2 \pm \sqrt{\left(\omega^2 + \omega_m^2 + 4g^2\right)^2 - 4\omega^2\omega_m^2}}
\end{equation}

\subsection{Modified}
Following ref. \citep{landauUSC}, where a prefactor $d$ is added before the diamagnetic term in Eq. \ref{eq:Hopfield2}, we tried to fit with the modified Hopfield model by varying this prefactor.
\begin{equation}
    \label{eq:Hopfield2}
    \omega_\pm = \dfrac{1}{\sqrt{2}}\sqrt{\omega_c^2 + 4dD\omega_c + \omega_c^2 \pm \sqrt{\left(\omega_c^2 + 4dD\omega_c - \omega_m^2\right)^2 + 16g^2\omega_c\omega_m}}
\end{equation}
where $D = g^2/\omega_m$

\begin{figure*}[h!]
    \includegraphics[width=\textwidth]{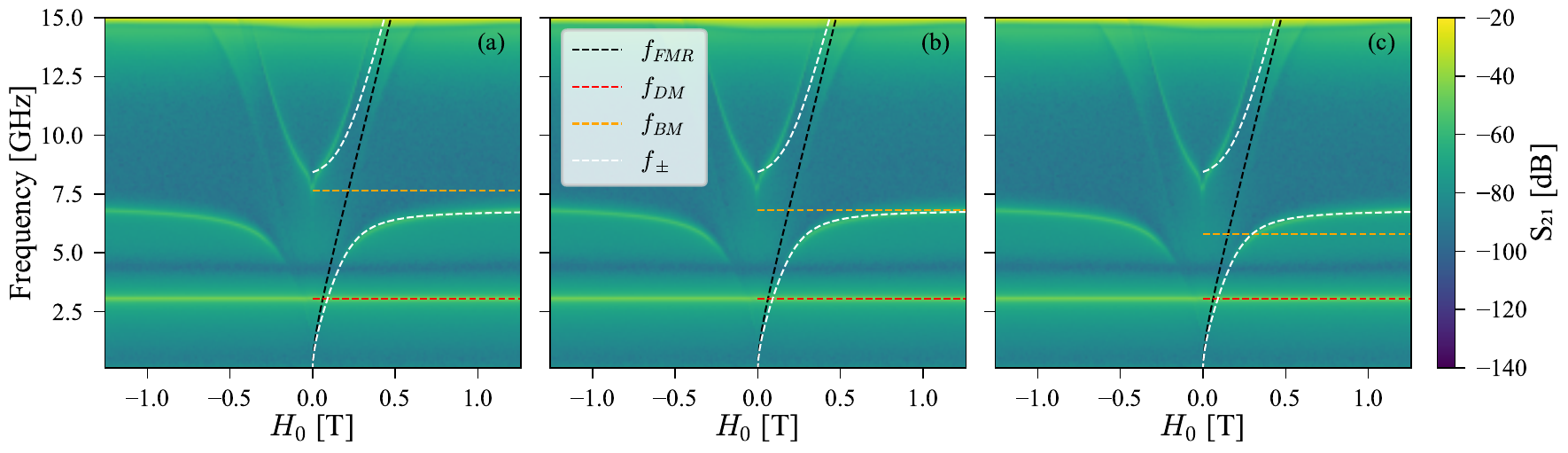}
    \captionsetup{width=\textwidth}
    \caption{ Transmission spectra versus the RF frequency and the $\mathbf{H}$-field. Fitted polariton branches are shown in white. The BM frequency (in orange) and the coupling strength are variables. The FMR is shown in black and the DM in red. Fits were be done with the modified Hopfield model where the prefactor is: (a) $d < 1$; (b) $d = 1$; and (c) $d > 1$.}
    \label{fig:Hopfield}
\end{figure*}
\figref{fig:Hopfield} shows the fit with the modified Hopfield model when the prefactor is less, equal, or more than 1 in respectively (a), (b), and (c). Let us notice that the standard Hopfield model is for $d=1$. Finally the only effect of this prefactor is equivalent to increase (for $d<1$) or decrease (for $d>1$) the BM frequency whereas it is needed to have a model which affects the FMR.

\newpage
\section{Measurements}
\label{appendix:measurements}
\subsection{CAV$_{01}$}
\begin{figure}[h!]
    \centering
    \includegraphics{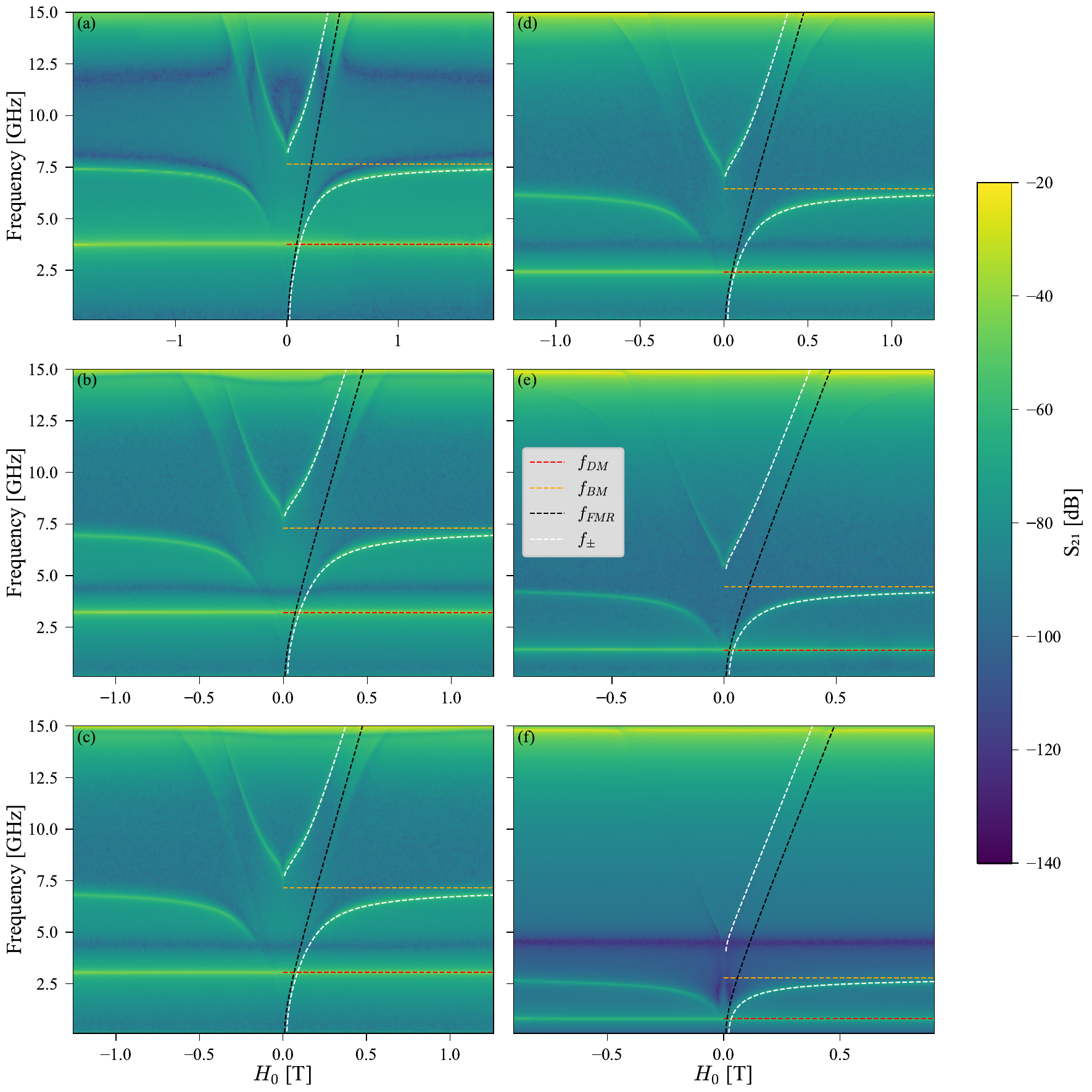}
    \caption{ Transmission spectra versus the RF frequency and the $\mathbf{H}$-field. Fitted polariton branches are shown in white. The BM frequency (in orange) and the coupling strength are variables. The FMR is shown in black and the DM in red. Fitted parameters are shown in Table \ref{tab:CAV01}.}
    \label{fig:CAV01}
\end{figure}
\begin{table}[h!]
    \caption{\label{tab:CAV01}%
        Cavity parameters from \figref{fig:CAV01}}
    \begin{ruledtabular}
        \begin{tabular}{cccccccccc}
            \textrm{Numbering}         &
            \textrm{$d$ [$\mu$m]}         &
            \textrm{$f_{DM}$ [GHz]} &
            \textrm{$f_{BM}$ [GHz]} &
            \textrm{$g/2\pi$ [GHz]}      &
            \textrm{$g/\omega$}      &
            \textrm{$g^2/2\pi\omega$ [GHz]}      &
            \textrm{$\Delta_m/2\pi$ [GHz]}  &
            \textrm{$f_{gap}$ [GHz]}                                         \\
            \colrule
            \textrm{(a)}     & 116 & 3.75   & 7.65 & 2.68 & 0.35 & 0.94 & 2.35 & 0.58\\
            \textrm{(b)}     & 75 & 3.19 & 7.31 & 2.62 & 0.36 & 0.94 & 2.29 & 0.54\\
            \textrm{(c)}     & 65  & 3.05  & 7.16   & 2.56 & 0.36 & 0.92 & 2.31 & 0.54\\
            \textrm{(d)}     & 36  & 2.40  & 6.44   & 2.41 & 0.37 & 0.90 & 2.27 & 0.64\\
            \textrm{(e)}     & 10  & 1.38  & 4.46   & 2.03 & 0.46 & 0.92 & 2.39 & 0.87\\
            \textrm{(f)}     & 3  & 0.81  & 2.80   & 1.64 & 0.59 & 0.96 & 2.59 & 1.22\\
        \end{tabular}
    \end{ruledtabular}
\end{table}

\subsection{CAV$_{02}$}
\begin{figure}[h!]
    \centering
    \includegraphics{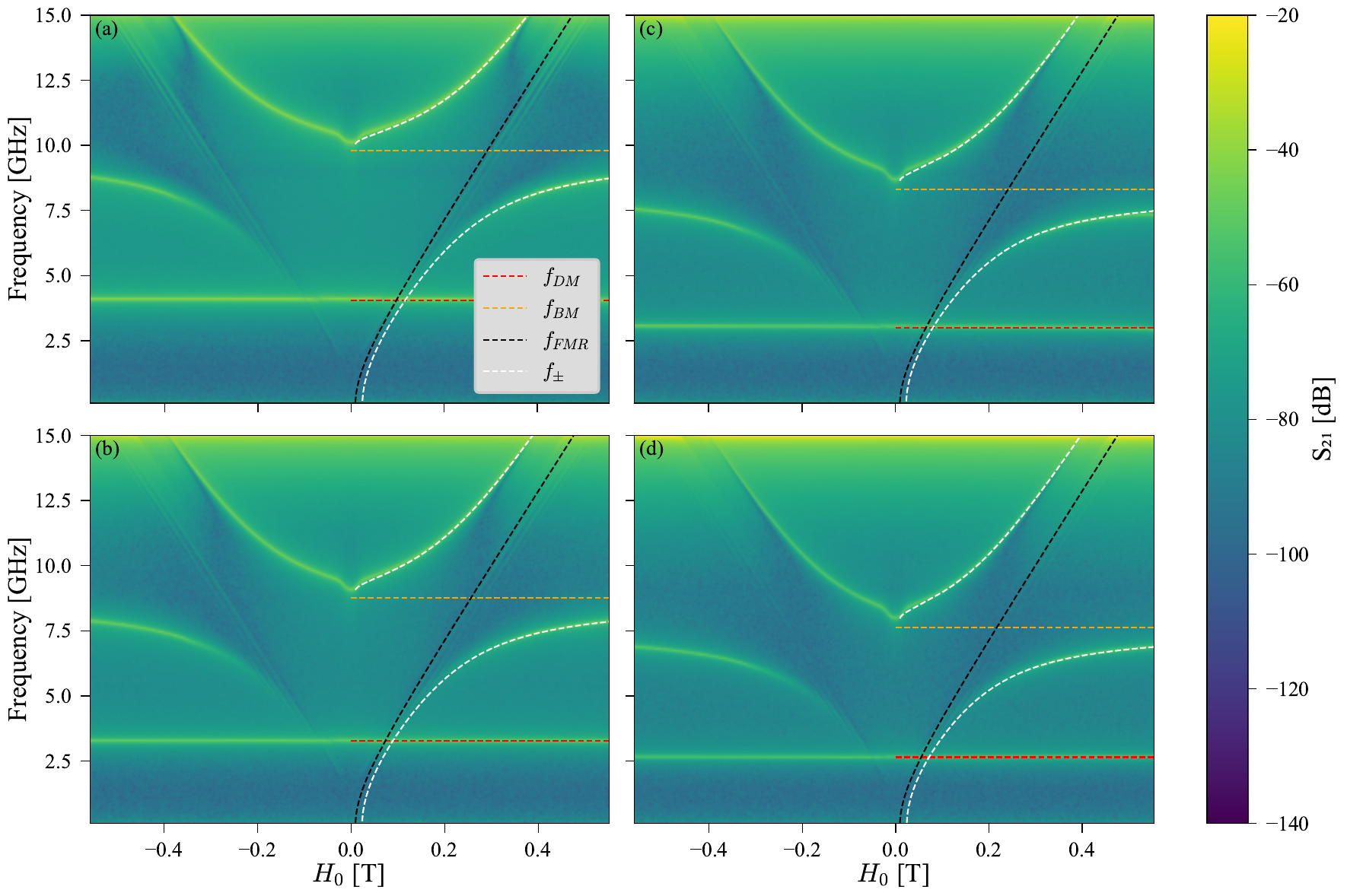}
    \caption{ Transmission spectra versus the RF frequency and the $\mathbf{H}$-field. Fitted polariton branches are shown in white. The BM frequency (in orange) and the coupling strength are variables. The FMR is shown in black and the DM in red. Fitted parameters are shown in Table \ref{tab:CAV02}.}
    \label{fig:CAV02}
\end{figure}
\begin{table}[h!]
    \caption{\label{tab:CAV02}%
        Cavity parameters from \figref{fig:CAV02}}
    \begin{ruledtabular}
        \begin{tabular}{cccccccc}
            \textrm{Numbering}         &
            \textrm{$f_{DM}$ [GHz]} &
            \textrm{$f_{BM}$ [GHz]} &
            \textrm{$g/2\pi$ [GHz]}      &
            \textrm{$g/\omega$}      &
            \textrm{$g^2/2\pi\omega$ [GHz]}      &
            \textrm{$\Delta_m/2\pi$ [GHz]}   &
            \textrm{$f_{gap}$ [GHz]}                                   \\
            \colrule
            \textrm{(a)}      & 4.06   & 9.79 & 2.72 & 0.28 & 0.76 & 1.63 & 0.24\\
            \textrm{(b)}      & 3.26 & 8.76 & 2.59 & 0.30 & 0.77 & 1.71 & 0.30\\
            \textrm{(c)}       & 3.01  & 8.32   & 2.52 & 0.30 & 0.76 & 1.74 & 0.31\\
            \textrm{(d)}       & 2.64  & 7.63   & 2.42 & 0.32 & 0.77 & 1.69 & 0.34\\
        \end{tabular}
    \end{ruledtabular}
\end{table}

\subsection{CAV$_{03}$}
\begin{figure}[h!]
    \centering
    \includegraphics{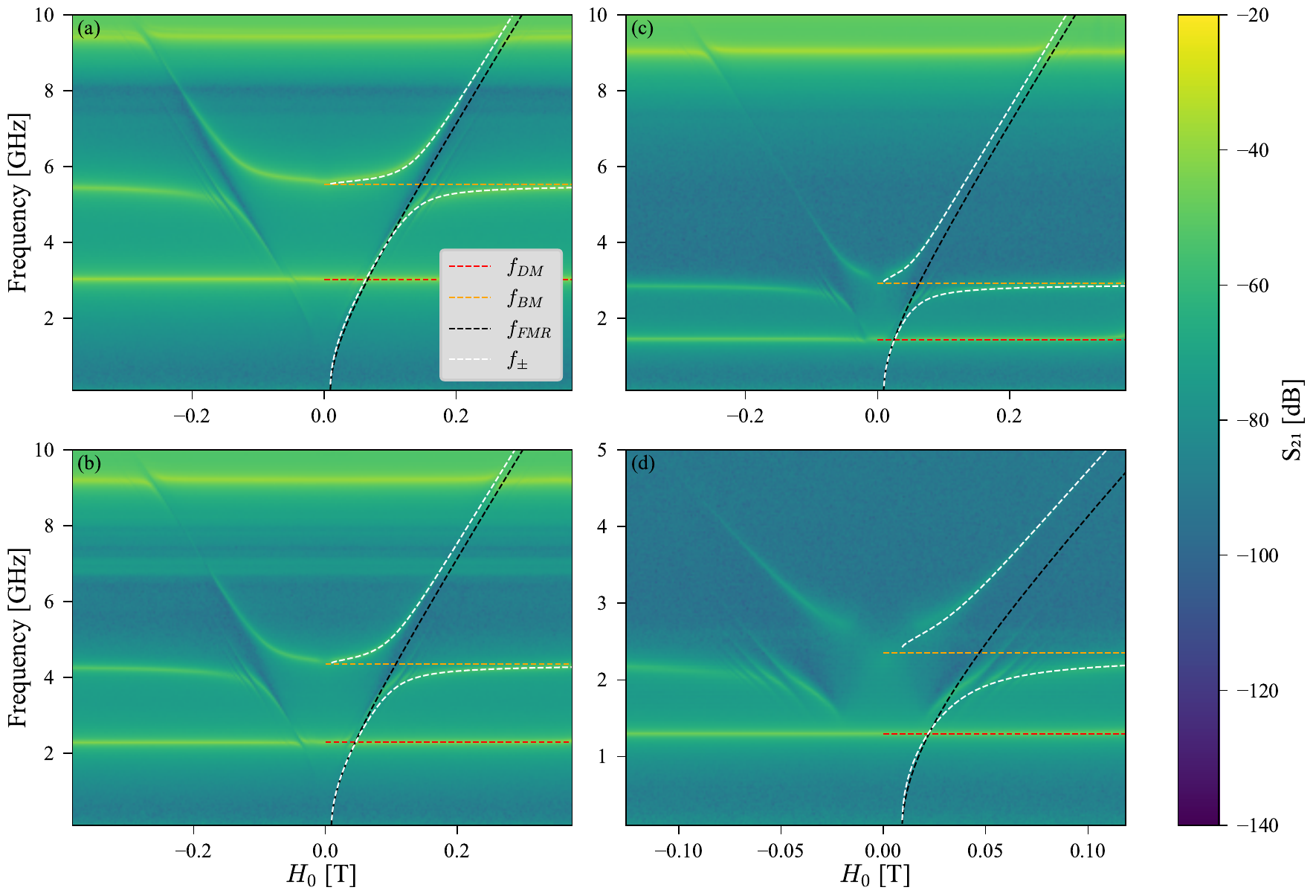}
    \caption{ Transmission spectra versus the RF frequency and the $\mathbf{H}$-field. Fitted polariton branches are shown in white. The BM frequency (in orange) and the coupling strength are variables. The FMR is shown in black and the DM in red. Fitted parameters are shown in Table \ref{tab:CAV03}.}
    \label{fig:CAV03}
\end{figure}
\begin{table}[h!]
    \caption{\label{tab:CAV03}Cavity parameters from \figref{fig:CAV03}}
    \begin{ruledtabular}
        \begin{tabular}{cccccccc}
            \textrm{Numbering}         &
            \textrm{$f_{DM}$ [GHz]} &
            \textrm{$f_{BM}$ [GHz]} &
            \textrm{$g/2\pi$ [GHz]}      &
            \textrm{$g/\omega$}      &
            \textrm{$g^2/2\pi\omega$ [GHz]}      &
            \textrm{$\Delta_m/2\pi$ [GHz]}   &
            \textrm{$f_{gap}$ [GHz]}                                   \\
            \colrule
            \textrm{(a)}      & 3.02   & 5.53 & 0.65  & 0.12 & 0.08 & 0.33 & 0.01\\
            \textrm{(b)}      & 2.29 & 4.36 & 0.69 & 0.16 & 0.11 & 0.29 & 0.02\\
            \textrm{(c)}       & 1.44  & 2.92   & 0.63 & 0.22 & 0.14 & 0.37 & 0.02\\
            \textrm{(d)}       & 1.30  & 2.35   & 0.58 & 0.25 & 0.14 & 0.50 & 0.05\\
        \end{tabular}
    \end{ruledtabular}
\end{table}

\newpage
\section{Gap Study}
\label{appendix:gap}
\begin{figure}[h!]
    \centering
    \includegraphics{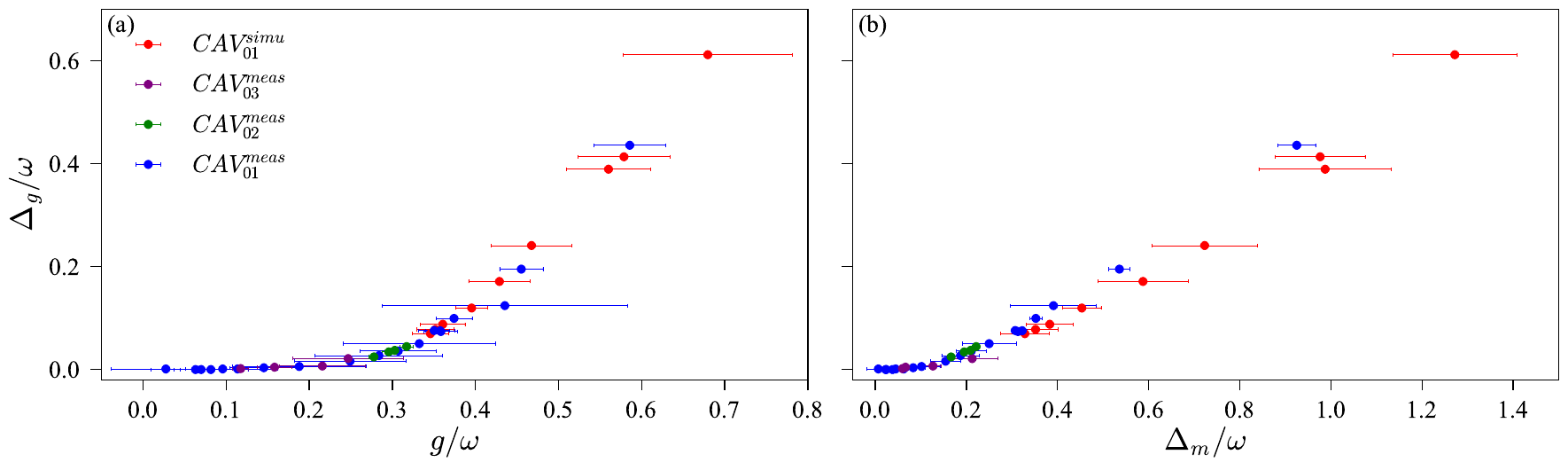}
    \caption{$\Delta_g/\omega$ versus (a) $g/\omega$; (b) $\Delta_m/\omega$.
    Shown are the FD simulations on CAV$_{01}$ in red and measurements in blue, in green for CAV$_{02}$, and in purple for CAV$_{03}$.}
    \label{fig:gap}
\end{figure}

Without adding $\Delta_m$ to the FMR in the Dicke model, and without applied static magnetic field, the frequency of the upper polariton is equal to the cavity one. However, when the FMR is shifted, an observable forbidden gap in frequency appears.
Considering \figref{fig:gap}, $\Delta_g/\omega$ is not observable when $g/\omega$ is equal or lower to 0.2. For higher $g/\omega$ values, $\Delta_g/\omega$ is quadratic, as shown in (a). In (b) is shown the evolution of $\Delta_g/\omega$ versus $\Delta_m/\omega$.

\end{document}